# A study of the spectrum resource leasing method based on ERC4907 extension

Zhiming Liang
*College of Electronics and Information Engineering*
*Shenzhen University*
Shenzhen, China
lzming0212@163.com

Bin Chen*
*College of Electronics and Information Engineering*
*Shenzhen University*
Shenzhen, China
bchen@szu.edu.cn
*Corresponding author

Litao Ye
*College of Electronics and Information Engineering*
*Shenzhen University*
Shenzhen, China
yelitao210708856@163.com

Chen Sun
*Wireless Network Research Department*
*Research and Development Center, Sony (China) Limited*
Beijing, China
chen.sun@sony.com

Shuo Wang
*Wireless Network Research Department*
*Research and Development Center, Sony (China) Limited*
Beijing, China
shuo.wang@sony.com

Zhe Peng
*Department of Industrial and Systems Engineering*
*The Hong Kong Polytechnic University*
Hong Kong, China
jeffrey-zhe.peng@polyu.edu.hk

*Abstract*—The ERC4907 standard enables rentable Non-Fungible Tokens (NFTs) but is limited to single-user, single-time-slot authorization, which severely limits its applicability and efficiency in decentralized multi-slot scheduling scenarios. To address this limitation, this paper proposes Multi-slot ERC4907 (M-ERC4907) extension method. The M-ERC4907 method introduces novel functionalities to support the batch configuration of multiple time slots and simultaneous authorization of multiple users, thereby effectively eliminating the rigid sequential authorization constraint of ERC4907. The experiment was conducted on the Remix development platform. Experimental results show that the M-ERC4907 method significantly reduces on-chain transactions and overall Gas consumption, leading to enhanced scalability and resource allocation efficiency.

## I. Introduction

The rapidly growing demand for wireless communication continually clashes with the inherent scarcity of spectrum resources. To address this crucial mismatch, Dynamic Spectrum Sharing (DSS) [1] has become an essential technology. DSS allows Secondary Users (SUs) to access spectrum resources that are temporarily idle or unused by Primary Users (PUs). By implementing this flexible access mechanism, DSS effectively minimizes resource waste caused by spectrum idleness, thereby significantly improving overall spectrum utilization efficiency[2]. However, traditional centralized DSS systems face several significant drawbacks, including reliance on trusted third parties, high operational expenses, and inherent vulnerability to manipulation[3]. These challenges ultimately limit their capacity to support efficient and secure large-scale spectrum allocation.

Blockchain [4][5] and smart contracts [6] provide a decentralized alternative by ensuring immutability, transparency, and automation[7], thereby reducing risks and enabling verifiable on-chain spectrum leasing. Existing studies have preliminarily explored blockchain-based decentralized DSS applications. In [8], an algorithm for decentralized operator transactions is proposed using consortium blockchain and game theory, which improves spectrum utilization efficiency without the involvement of a third party. The study primarily focuses on improving spectrum utilization and increasing operators' revenues, while neglecting the optimization of spectrum allocation efficiency. In [9], blockchain is utilized as a distributed database to enhance the security and efficiency of dynamic spectrum access in mobile cognitive radio networks. It adopts an auction mechanism based on a first-come-first-served queue, which fails to meet the demand for efficient allocation of the same spectrum resource across multiple different time slots simultaneously. In [10], a Multi-operators Spectrum Sharing smart contract is proposed, which realizes the trusted allocation and settlement of spectrum resources. However, the proposed multi-operator spectrum auction and free trading mechanisms involve relatively complex sharing procedures, resulting in excessive on-chain operations and reduced spectrum allocation efficiency. None of the aforementioned studies discuss how to maintain the uniqueness of spectrum resources after sharing. The lack of a clear definition for spectrum uniqueness may lead to spectrum access conflicts. To address this issue, most recent studies have introduced NFTs to build on-chain verifiable access licenses for spectrum resources[11][12]. In [13], spectrum tokens are constructed based on

ERC721 [14] standard. Although the unique identification of spectrum is realized, the problem of how to return spectrum tokens arises because spectrum tokens ownership is transferred along with the lease. Recent work [15] applies the ERC4907 [16] rentable NFT standard to DSS. ERC4907 can separate the ownership and usage rights of NFTs[17], thereby overcoming the limitations of ERC721. Based on this standard, the authors mint Non-Fungible Spectrum Tokens (NFSTs) to demonstrate that decentralized leasing can improve the efficiency and security of spectrum sharing. However, ERC4907 only supports the authorization of user usage rights for a single time slot, which means that a user must wait until the current lease expires before renting the next time slot. This sequential authorization limitation highlights the lack of batch and parallel authorization capability, which significantly restricts its applicability in multi-slot spectrum allocation scenarios.

To overcome these limitations, we propose the M-ERC4907 method, an extension method of ERC4907 that enables batch configuration of idle time slots and simultaneous batch authorization of multiple users. By breaking the sequential authorization dependency of ERC4907, the M-ERC4907 method reduces redundant transactions, lowers Gas consumption, and achieves efficient, scalable spectrum allocation in DSS scenarios.

## II. SYSTEM MODEL

The decentralized DSS model is shown in Fig. 1. It includes six core components: (a) Blockchain, which stores spectrum metadata and auction records and executes smart contracts; (b) Spectrum Management Authority, which regulates spectrum allocation and compliance; (c) PU, the licensed spectrum owner and contract deployer; (d) SU, the demander of spectrum resources; (e) Tokenization Module, which converts spectrum resources into Non-Fungible Spectrum Tokens (NFSTs); and (f) Auction Module, which manages leasing, usage authorization, and automatic reclamation. These elements together enable a decentralized, transparent, and verifiable DSS system.

After receiving authorization from the Spectrum Management Authority, the PU leases idle spectrum through the DSS system. The process is shown in Fig. 2, with the following steps:

1) The PU uses the Spectrum Resource Tokenization Module to mint the spectrum it holds into corresponding NFSTs based on the ERC4907 standard. Each token is minted with metadata including the spectrum's geographic location, frequency band, and owner address.

2) The M-ERC4907 method allows the PU to pre-configure multiple idle time slots on the same NFST. After the PU schedules multiple idle time slots in advance for the NFST it holds, the SU can invoke the smart contract to query the available idle spectrum information of NFSTs at its current location.

3) After the PU initiates the auction, the SU can invoke the smart contract to submit standardized bidding information to participate in the auction.

4) Upon auction conclusion, the smart contract allocates the auction proceeds according to predefined rules and automatically sets the user of each idle time slot of the NFST to the highest bidder for that specific time slot. The usage start and end times correspond to the start and end times of the idle time slot selected during the auction. During the lease period, the highest bidder is allowed to use the corresponding spectrum resources in accordance with the protocol.

5) When the lease term expires, the contract resets the user address of the time slot to the zero address, thereby releasing the spectrum usage right for the user of the next time slot.

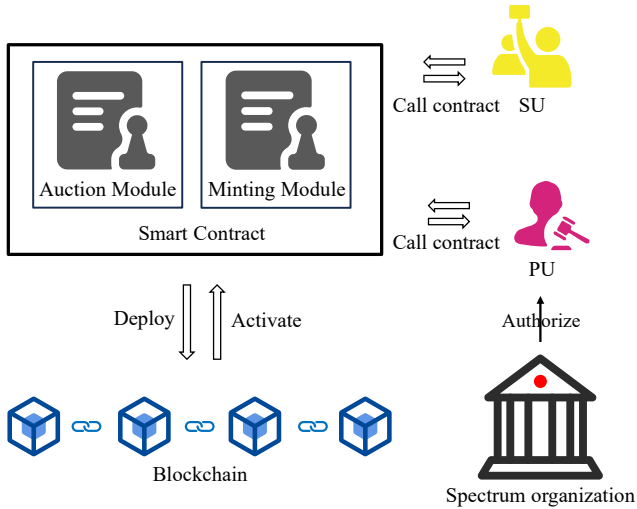

Fig. 1. System model.

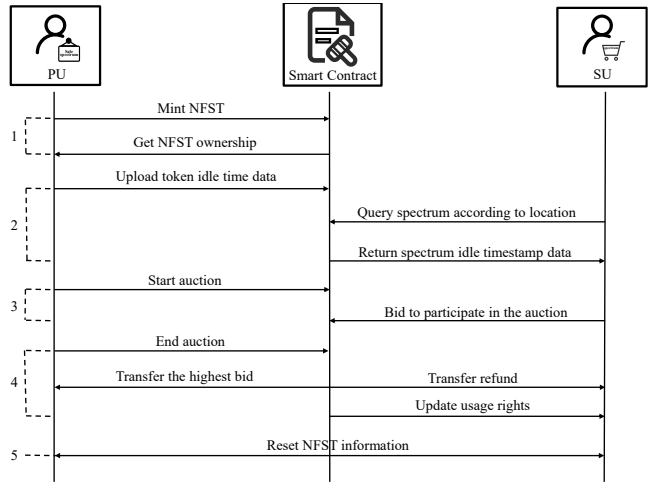

Fig. 2. DSS process.

Accordingly, the M-ERC4907 method provides new functionalities absent in ERC4907, namely multi-slot batch configuration and simultaneous authorization. This overcomes the sequential authorization limitation of ERC4907, enabling more efficient spectrum leasing in multi-user and multi-slot scenarios.

## III. SMART CONTRACT DESIGN

We design a smart contract that extends ERC4907 to implement the proposed the M-ERC4907 method for decentralized DSS. The key novelty lies in supporting multi-slot configuration and simultaneous multi-user authorization, which are not available in ERC4907.

### A. Spectrum Resource Tokenization and Idle Time Configuration

Following the scheme in [15], spectrum resources are minted into NFSTs using the ERC4907 standard. Each NFST includes metadata such as the spectrum owner, frequency band, and geographic location. As shown in Fig. 3, to extend this model, we introduce a new function, setSpectrumIdleTime, which enables the configuration of multiple non-overlapping time slots for each NFST. These slots are embedded in the metadata and serve as the basis for leasing spectrum resources on a per-slot basis.

To address the limitation of ERC4907 that only supports configuring one time slot at a time and lacks the capability to configure multiple time slots simultaneously, the M-ERC4907 method enables the PU to predefine multiple non-overlapping idle time slots for the subsequent simultaneous authorization of different users. This mechanism significantly enhances the flexibility of spectrum resource configuration across time slots while effectively reducing the number of on-chain operations and associated costs.

```
Function setSpectrumIdleTime
  Input: tokenId, idleStartTimestamps, idleEndTimestamps
    if (Authorized[caller]=true) then
      NFSTMetadata[tokenId].idleStartTimestamps ← idleStartTimestamps
      NFSTMetadata[tokenId].idleEndTimestamps ← idleEndTimestamps
      NFSTMetadata[tokenId].idleTimeSlotNum ← idleEndTimestamps.length
      return true
    else return false
End Function
```

Fig. 3. Batch configuration of spectrum idle times.

### B. Batch Authorization of Spectrum Users

In ERC4907, the function setUser authorizes only one user for one lease period. In Fig. 4, the M-ERC4907 method introduces the batchSetUser function, which maps multiple users to their respective time slots simultaneously. This eliminates the sequential authorization dependency of ERC4907, where a user must wait until the previous lease expires, and ensures that all slots can be allocated in parallel.

```
Function batchSetUser
  Input: tokenId, users, startTimes, endTimes
      batchUsers[tokenId].users ← users
      batchUsers[tokenId].startTimes ← startTimes
      batchUsers[tokenId].endTimes ← endTimes
End Function
```

Fig. 4. Batch authorization of spectrum users.

## C. Auction Mechanism for DSS

The M-ERC4907 method integrates a multi-slot synchronous English Auction, in which bidders compete for several time slots simultaneously. The auction process is implemented through three core functions. The startAuction function shown in Fig. 5 initializes the auction by setting parameters such as NFST ID, auction duration, starting price, and available idle time slots. The Bid function shown in Fig. 6 allows SUs to place ascending bids for one or multiple time slots, and the contract records and updates the current highest bid for each slot. Finally, the endAuction function shown in Fig. 7 finalizes the auction, allocates all slots in batch to the highest bidders, guarantees exclusive usage during the lease, and resets the slots automatically upon expiration.

This English Auction design ensures transparency, prevents conflicts, and enables efficient execution of simultaneous multi-slot spectrum allocation.

```
Function startAuction
  Input: tokenID, auctionDuration, timeSlotCount, bottomPrice
    if(Authorized[caller]=true and currentUserOf(tokenID)=address(0)) then
        auctions[tokenId].auctionEndTime ← now + auctionDuration
        auctions[tokenId].hasEnded ← false
        for (i = 0; i < timeSlotCount; i++)
            timeSlotsByToken[tokenId][i].highestBidder ← address(0)
            timeSlotsByToken[tokenId][i].highestBid ← 0
            timeSlotsByToken[tokenId][i].isFinalized ← false
            timeSlotsByToken[tokenId][i].bottomPrice ← bottomPrice
        return true
    else return false
End Function
```

Fig. 5. Auction initialization.

```
Function Bid
  Input: tokenId, timeSlotId
    if(Authorized[caller]=true or caller=highestBidder) then
        return false
    if(bid + callerRefund > highestBid and now < auctionEndTime) then
        timeSlotsByToken[tokenId][timeSlotId].highestBid ← bid + callerRefund
        timeSlotsByToken[tokenId][timeSlotId].highestBidder ← caller
        biddersByTimeSlot[tokenId][timeSlotId].push(caller)
        return true
    else return false
End Function
```

Fig. 6. Auction bidding.

```
Function endAuction
  Input: tokenId
    if(Authorized[caller]=true and now > auctionEndTime) then
      for (i = 0; i < auctions[tokenId].timeSlotCount; i++)
        highestBidder ← timeSlotsByToken[tokenId][i].highestBidder
        highestBid ← timeSlotsByToken[tokenId][i].highestBid
        winners[i] ← highestBidder
        timeSlotsByToken[tokenId][i].highestBid ← 0
        payable(seller).transfer(highestBid)
        for (j = 0; j < biddersByTimeSlot[tokenId][i].length; j++)
          bidder ← biddersByTimeSlot[tokenId][i][j]
          if (bidder ≠ highestBidder) then
            bidderRefund ← refundBalances[bidder][tokenId][i]
            refundBalances[bidder][tokenId][i] ← 0
            payable(bidder).transfer(bidderRefund)
      batchSetUser(tokenId, winners, startTimes, endTimes)
      auctions[tokenId].hasEnded = true
      return true
    else return false
End Function
```

Fig. 7. Auction termination.

## IV. EXPERIMENTAL RESULTS AND ANALYSIS

### A. Experimental Procedure

To validate the effectiveness of the M-ERC4907 method in enabling batch authorization of multiple time slots and supporting dynamic spectrum allocation, we deployed the smart contracts on the Ethereum London test network using the Remix VM development platform. Prior to the start of the experiment, one PU and six SUs were selected to participate.

The PU mints its authorized spectrum resources into NFST. As shown in Fig. 8, where PU is the owner of the NFST with ID 1 and its address is 0xdD8...92148.

```
"from": "0xD1ee42fdA217994CF17F7D37E8909FA2c30Ca192",
"topic": "0xf4283a178641472779acc1a10cba34ef7277e3de9c3f0dc5d3461e17e24c2483",
"event": "SpectrumTokenization",
"args": {
    "0": "1",
    "1": "0xdD870fA1b7C4700F2BD7f44238821C26f7392148",
    "2": "600MHz",
    "3": "800MHz",
    "4": "CityA",
    "tokenId": "1",
    "owner": "0xdD870fA1b7C4700F2BD7f44238821C26f7392148",
    "startFreq": "600MHz",
    "endFreq": "800MHz",
    "location": "CityA"
}
```

Fig. 8. NFST minting result.

As shown in Fig. 9, PU uses the setSpectrumIdleTime function to batch-set 3 idle time slots for the NFST with ID 1. Among them, time slot 0 spans from 1749476800 to 1749476900, time slot 1 spans from 1749477000 to 1749477100, and time slot 2 spans from 1749477200 to 1749477300, where all times are represented using Unix timestamps.

```
"from": "0xD1ee42fdA217994CF17F7D37E8909FA2c30Ca192",
"topic": "0x0f089b14dea601ddfd2fea84663c5edacdf86e64a28057d5886806e48eca7e76",
"event": "SetNFSTIdleTime",
"args": {
        "0": "1",
        "1": [
                "1749476800",
                "1749477000",
                "1749477200"
        ],
        "2": [
                "1749476900",
                "1749477100",
                "1749477300"
        ],
        "3": "3",
        "tokenId": "1",
        "newSpectrumIdleStartTimes": [
                "1749476800",
                "1749477000",
                "1749477200"
        ],
        "newSpectrumIdleEndTimes": [
                "1749476900",
                "1749477100",
                "1749477300"
        ],
        "idleTimeSlot": "3"
}
```

Fig. 9. Result of idle time slot configuration.

PU sets the auction duration and the starting price for each idle time slot for the NFST with ID 1 to initialize the auction. Fig. 10 presents the on-chain status after the auction is initialized.

```
"from": "0xD1ee42fdA217994CF17F7D37E8909FA2c30Ca192",
"topic": "0x5862ab4ff3b3fa555eee0ff8011278dd35b8834552b7a2bb96b4d06e5e915730",
"event": "auctionStart",
"args": [
        "0": "1",
        "1": "180",
        "2": "3",
        "3": "10000000000000000000",
        "4": "Auction in progress",
        "tokenId": "1",
        "auctionDuration": "180",
        "timeSlotCount": "3",
        "timeSlotBottomPrice": "10000000000000000000",
        "auctionStatus": "Auction in progress"
}
```

Fig. 10. Result of auction initialization.

After the auction starts, SUs can participate in bidding in parallel for the usage rights of the 3 time slots of the NFST with ID 1. The bidding status for each time slot is shown in Table I. For time slot 1, SU1 first bids 11 ETH, followed by SU2 and SU3 participating in the bidding with bid amounts of 12 ETH and 13 ETH respectively, and finally SU1 increases the bid by 5 ETH. For time slot 2, SU4 first bids 11 ETH, then SU5 participates in the bidding with a bid amount of 13 ETH, and finally SU4 increases the bid by 3 ETH. For time slot 3, only SU6 bids 15 ETH.

TABLE I. BIDDING SITUATION WITH DIFFERENT SLOTS

| time slot 0 | | | |
|---|---|---|---|
| bid order | bidders | address | bid amount (ETH) |
| 1 | SU1 | 0x5B3…eddC4 | 11 |
| 2 | SU2 | 0xAb8…35cb2 | 12 |
| 3 | SU3 | 0x4B2…C02db | 13 |
| 4 | SU1 | 0x5B3…eddC4 | 5 |
| time slot 1 | | | |
| bid order | bidders | address | bid amount (ETH) |
| 1 | SU4 | 0x787...cabaB | 11 |
| 2 | SU5 | 0x617...5E7f2 | 12 |
| 3 | SU4 | 0x787...cabaB | 3 |
| time slot 2 | | | |
| bid order | bidders | address | bid amount (ETH) |
| 1 | SU6 | 0x17F...8c372 | 11 |

After the auction ends, PU ends the auction. In Fig. 11, it shows the results of batch authorization of user usage rights after the auction ends. The contract designates winning bidders SU1, SU4, and SU6 as the users of time slots 0, 1, and 2 of the NFST, and their lease periods are from 1749476800 to 1749476900, 1749477000 to 1749477100, and 1749477200 to 1749477300, where all times are represented using Unix timestamps.

```
"from": "0xD1ee42fdA217994CF17F7D37E8909FA2c30Ca192",
"topic": "0xfb4bad7df068762e64d9213a3b032bd620fb483d61b473f4cb2203fcd740e18e",
"event": "UpdateUser",
"args": {
    "0": "1",
    "1": [
        "0x5B38Da6a701c568545dCfcB03FcB875f56beddC4",
        "0x78731D3Ca6b7E34aC0F824c42a7cC18A495cabaB",
        "0x17F6AD8Ef982297579C203069C1DbfFE4348c372"
    ],
    "2": [
        "1749476800",
        "1749477000",
        "1749477200"
    ],
    "3": [
        "1749476900",
        "1749477100",
        "1749477300"
    ],
    "tokenId": "1",
    "users": [
        "0x5B38Da6a701c568545dCfcB03FcB875f56beddC4",
        "0x78731D3Ca6b7E34aC0F824c42a7cC18A495cabaB",
        "0x17F6AD8Ef982297579C203069C1DbfFE4348c372"
    ],
    "startUseTimes": [
        "1749476800",
        "1749477000",
        "1749477200"
    ],
    "endUseTimes": [
        "1749476900",
        "1749477100",
        "1749477300"
    ]
}
```

Fig. 11. Result of batch authorization of user usage rights.

## B. Performance Analysis

We further designed and conducted comparative experiments. User usage right authorization operations were performed on the same NFST across 10 different time slots, and the gas consumption and transaction counts of ERC4907 and the M-ERC4907 method were tested respectively to evaluate the performance of the two methods under different numbers of time slots.

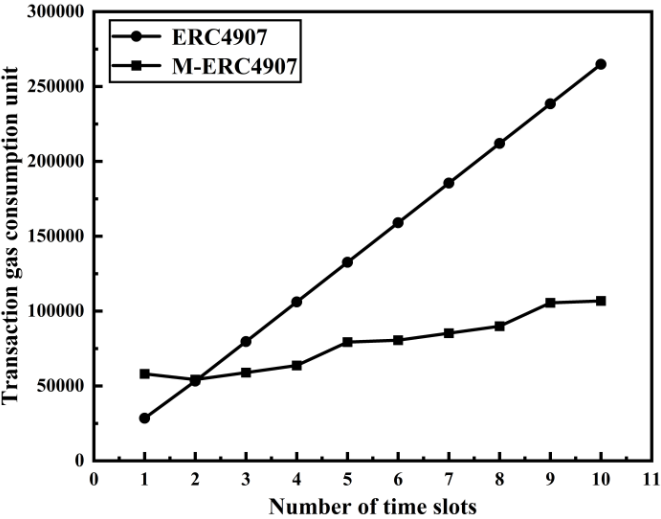

Fig. 12. Comparison of Gas consumption per transaction.

To evaluate the performance improvements of the M-ERC4907 method in terms of on-chain cost, this paper conducts comparative experiments on Gas consumption between ERC4907 and the M-ERC4907 method under user authorization scales with different time slots. The user authorization operations for each number of time slots are repeated ten times, and the average Gas consumption value is adopted to ensure the reliability of the results. The Gas consumption results are shown in Fig. 12. When authorizing a single time slot, the Gas consumption of the M-ERC4907 method is higher than that of ERC4907. As the number of time slots increases to two, the difference in Gas consumption between the two methods gradually narrows, but the M-ERC4907 method is still slightly higher. When the number of time slots exceeds two, the Gas consumption of ERC4907 shows an obvious linear growth trend, while the growth rate of the Gas consumption of the M-ERC4907 method is relatively slow. Overall, as the number of time slots increases, the average growth in gas consumption of ERC4907 is 26,270 Gas units for each additional time slot's user usage right authorization. In contrast, the average growth of the M-ERC4907 method is approximately 5,409 Gas units. The experimental results indicate that when only one or two time slots of usage rights need to be authorized, the batch authorization mechanism of the M-ERC4907 method has not fully demonstrated its performance advantages. The main reason is that the M-ERC4907 method requires initializing the array storage structure when initiating the batch authorization process. Fig. 8 presents the three arrays that need initialization in the batch authorization process. In addition, batch authorization also involves multiple array writing operations. These additional overheads offset the potential performance gains in scenarios with a small number of time slots. However, as the number of time slots further increases, ERC4907 exhibits linear growth in Gas consumption because each time slot requires an independent transaction. In contrast, the M-ERC4907 method grows more slowly due to its batch authorization mechanism. Minor fluctuations in the M-ERC4907 method trend are caused by the Ethereum Virtual Machine's 32-byte storage mechanism. When the number of bytes stored in an array exceeds 32, a new storage slot will be allocated, leading to a sudden increase in Gas consumption. However, this does not affect the overall scalability advantage.

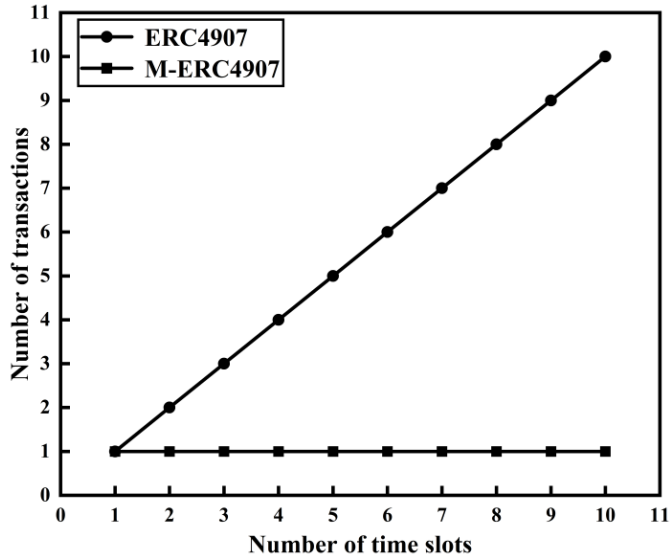

Fig. 13. Comparison of on-chain transaction operation counts.

In addition to Gas consumption, the number of required transactions was also measured, as shown in Fig. 13. The results indicate that the transaction count of ERC4907 exhibits linear growth with the increase in the number of time slots, as the authorization of usage rights to users for each time slot must be implemented through an independent transaction. In contrast, the M-ERC4907 method supports the simultaneous authorization of usage rights to users for multiple non-overlapping time slots in a single transaction. Therefore, when the Gas consumption of a single transaction does not exceed the maximum on-chain Gas limit, the authorization process can be completed in one transaction regardless of how the number of time slots increases. This substantially reduces the overall number of transactions, alleviates potential network congestion, and enhances scalability in large-scale DSS scenarios.

## V. CONCLUSION

This paper proposes the M-ERC4907 method, an extension of ERC4907 that supports multi-time-slot batch configuration and parallel user authorization, thereby overcoming the limitation of ERC4907 that only sequential processing is possible for NFST usage right authorization. The decentralized DSS system built based on this method has been experimentally verified to achieve efficient parallel spectrum resource allocation in multi-user and multi-time-slot scenarios. In terms of performance, comparisons with ERC4907 highlight the advantages of the M-ERC4907 method. As the number of time slots increases, due to batch processing, the gas consumption grows at a slower rate, and the number of on-chain transactions remains significantly reduced. These improvements enhance scalability, increase efficiency by minimizing redundant operations, and reduce overall gas costs. Therefore, the M-ERC4907 method achieves excellent scalability in multi-slot scenarios while maintaining the security and verifiability of ERC4907.

ACKNOWLEDGMENT

Guangdong Province Graduate Education Innovation Program Project[2024JGXM_163], Shenzhen University Graduate Education Reform Research Project [SZUGS2023JG02], Foundation of Shenzhen [20220810142731001]. Shenzhen University

## REFERENCES

[1] B. Wang and K. J. R. Liu, "Advances in cognitive radio networks: A survey," in IEEE Journal of Selected Topics in Signal Processing, vol. 5, no. 1, pp. 5-23, Feb. 2011.

[2] S. T. Muntaha, P. I. Lazaridis, M. Hafeez, Q. Z. Ahmed, F. A. Khan, and Z. D. Zaharis, "Blockchain for dynamic spectrum access and network slicing: A review," IEEE Access, vol. 11, pp. 17922–17944, 2023.

[3] L. Perera, P. Ranaweera, S. Kusaladharma, S. Wang and M. Liyanage, "A Survey on Blockchain for Dynamic Spectrum Sharing," in IEEE Open Journal of the Communications Society, vol. 5, pp. 1753-1802, 2024.

[4] J. Golosova and A. Romanovs, "The Advantages and Disadvantages of the Blockchain Technology," 2018 IEEE 6th Workshop on Advances in Information, Electronic and Electrical Engineering (AIEEE), Vilnius, Lithuania, 2018, pp. 1-6.

[5] Z. Peng et al., "VFChain: Enabling Verifiable and Auditable Federated Learning via Blockchain Systems," in IEEE Transactions on Network Science and Engineering, vol. 9, no. 1, pp. 173-186, 1 Jan.-Feb. 2022.

[6] M. Kumarathunga, R. Neves Calheiros, and A. Ginige, "Sustainable microfinance outreach for farmers with blockchain cryptocurrency and smart contracts," Int. J. Comput. Theory Eng., pp. 9–14, 2022.

[7] L. Ye, B. Chen, C. Sun, S. Wang, P. Zhang and S. Zhang, "A Study of Semi-Fungible Token based Wi-Fi Access Control," 2024 IEEE 24th International Conference on Communication Technology (ICCT), Chengdu, China, 2024, pp. 722-727.

[8] S. Han and X. Zhu, "Blockchain Based Spectrum Sharing Algorithm," 2019 IEEE 19th International Conference on Communication Technology (ICCT), Xi'an, China, 2019, pp. 936-940.

[9] K. Kotobi and S. G. Bilen, "Secure Blockchains for Dynamic Spectrum Access: A Decentralized Database in Moving Cognitive Radio Networks Enhances Security and User Access," in IEEE Vehicular Technology Magazine, vol. 13, no. 1, pp. 32-39, March 2018.

[10] S. Zheng, T. Han, Y. Jiang and X. Ge, "Smart Contract-Based Spectrum Sharing Transactions for Multi-Operators Wireless Communication Networks," in IEEE Access, vol. 8, pp. 88547-88557, 2020.

[11] X. Shao, P. Cao, S. Wang, W. Wang, B. Zhou and C. Sun, "Non-Fungible Token Enabled Spectrum Sharing for 6G Wireless Networks," 2023 IEEE Globecom Workshops (GC Wkshps), Kuala Lumpur, Malaysia, 2023, pp. 1075-1080.

[12] Z. Zhou, B. Chen, C. Sun, P. Zhang and S. Wang, "Leveraging NFTs for Spectrum Securitization in 6G Networks," 2025 5th International Conference on Computer Science and Blockchain (CCSB), Shenzhen, China, 2025, pp. 172-177.

[13] T. Ariyarathna, P. Harankahadeniya, S. Isthikar, N. Pathirana, H. M. N. D. Bandara and A. Madanayake, "Dynamic Spectrum Access via Smart Contracts on Blockchain," 2019 IEEE Wireless Communications and Networking Conference (WCNC), Marrakesh, Morocco, 2019, pp. 1-6.

[14] W. Entriken, D. Shirley, J. Evans, and N. Sachs, "ERC-721: Non-Fungible Token Standard," Ethereum Improvement Proposals, no. 721, Jan. 2018. [Online]. Available: https://eips.ethereum.org/EIPS/eip-721.

[15] L. Ye et al., "Dynamic Spectrum Sharing Based on the Rentable NFT Standard ERC4907," 2024 13th International Conference on Communications, Circuits and Systems (ICCCAS), Xiamen, China, 2024, pp. 503-507.

[16] L. Anders and Shrug, "ERC-4907: Rental NFT, an Extension of EIP-721," Ethereum Improvement Proposals, no. 4907, Mar. 2022. [Online]. Available: https://eips.ethereum.org/EIPS/eip-4907.

[17] Z. Lin, B. Chen, Y. Chen, P. Zhang and H. Wang, "A Study of Decentralized Identifiers-Based Campus Public Resources Sharing Application," 2024 4th International Conference on Computer Science and Blockchain (CCSB), Shenzhen, China, 2024, pp. 362-366.